# Characteristics of X-point Lobe Structures in Single-Null Discharges on MAST

J. R. Harrison, A. Kirk, I. T. Chapman, P. Cahyna[2], Yueqiang Liu, E. Nardon[3], A. J. Thornton

EURATOM/CCFE Fusion Association, Culham Science Centre, Abingdon, Oxon OX14 3DB, UK
[2]Institute of Plasma Physics AS CR v.v.i., Association EURATOM/IPP.CR, Prague, Czech Republic
[3]Association Euratom/CEA, CEA Cadarache, F-13108, St. Paul-lez-Durance, France

**Abstract**

Lobe structures due to the application of resonant magnetic perturbations (RMPs) have been observed using wide-angle imaging of light from $He^{1+}$ ions in the vicinity of the lower X-point in MAST. The data presented are from lower single-null discharges where RMPs of toroidal mode number, n, of 4 and 6 were applied. It has been found that, above a threshold value, the lobe structures extend radially, linearly with the coil current, both in L-mode and H-mode. It is observed that after the application of the RMP, as the toroidal rotation in the confined plasma decreases, the lobes extend radially, suggesting the plasma is less effectively screening the RMP field. Comparing the imaging data with results from vacuum modelling shows that this technique can accurately predict the number and poloidal location of the lobes, but over-estimates their radial extent. More accurate estimates of the extent of the lobes can be made by accounting for plasma screening of the RMP field. Qualitative agreement between simulation and experiment is found if it is assumed that the RMP penetrates 2% in normalised radius from the last closed flux surface.

# 1. Introduction

The operation of next-generation of fusion devices with substantial auxiliary heating requires a means of controlling large edge localised modes (ELMs). This is required in order to protect plasma-facing surfaces on the first wall and divertor from the transient heat loads, predicted to be 20MJm$^{-2}$ in ITER [1]. However, experimental data suggests that in order to minimise damage to tungsten plasma-facing components, in the form of cracking and melting of surfaces, the maximum tolerable heat load is ~1 MJm$^{-2}$ [2]. The application of resonant magnetic perturbations (RMPs) is widely used as a technique for controlling ELMs in several fusion devices such as DIII-D [3], JET [4], ASDEX-U [5], MAST [6] and KSTAR [7], although their effects on the plasma are not fully understood. Studies on the aforementioned fusion devices have found that RMPs induce a reduction in particle density in the confined plasma (so-called density "pump-out"), strike point splitting [8], braking of toroidal rotation [9], appearance of lobe structures at the X-point [10], corrugation of the plasma edge [11] an increase in ELM frequency or the complete suppression of ELMs.

Studies of the effects of RMPs on the plasma magnetic topology have been carried out on many devices, and have mostly concentrated on measurements of split divertor heat flux patterns. In more recent studies, experimental evidence has been found for the modification of the magnetic topology around the X-point, i.e. the formation of lobe structures, as seen on MAST [10] and DIII-D [12]. These lobe structures were predicted theoretically [3, 8] before they were found experimentally, and coincide with intersecting stable and unstable manifolds due to the non-axisymmetric RMP field. The implications of these lobe structures are an area of on-going research, such as the link between the formation of X-point lobe structures and ELM mitigation [13].

In this paper, visible light imaging of X-point lobe structures due to RMPs with toroidal mode number, n, of 4 and 6 is presented. In section 2, analysed data are presented, showing the dependence of radial lobe extent with RMP current and indications of the role of plasma screening in setting the lobe extent. In section 3, the data is compared with results from vacuum modelling calculations, highlighting the strengths and shortcomings of the vacuum modelling technique to predict the size of the lobes.

## 1.1 Magnetic Structure of Lobes on MAST

MAST is fitted with 18 internal coils for the purposes of ELM control, with 12 mounted below the mid-plane and 6 above. The lower single null plasmas investigated in this study are a significant distance from the upper row of ELM coils, so only the lower row of coils is used. The set of RMP coils allows perturbations with toroidal mode number, n, of n= 3, 4, 6 using coils located below the mid-plane. This highly flexible set of coils can be used to produce a wide variety of RMP fields, each giving rise to different changes in the magnetic topology near the X-point. These changes can be predicted to varying degrees of sophistication using vacuum modelling codes such as ERGOS [14], which can include an ad-hoc model of the plasma response to the applied RMP [15] or MHD simulations of the plasma response using sophisticated codes such as MARS-F [9]. Figure 1 shows the results of ERGOS vacuum modelling calculations based on one magnetic equilibrium with n = 3, 4, 6 RMP fields superposed with the equilibrium fields. As the applied RMP fields are non-axisymmetric, the poloidal location of the lobe structures vary with the toroidal location field lines start from.

## 2. Imaging of X-point Lobes

The shape of the plasma boundary in the vicinity of the X-point is monitored using a filtered CCD camera with a tangential view of the lower divertor. The camera viewing geometry is shown schematically in Figure 2. Due to the position of the poloidal field coils inside MAST, it is only possible to view the X-point when the plasma is shifted downward into a single-null configuration, as the X-point is obscured during double-null operation. Consequently, all of the data presented in this paper are from single-null discharges. The data taken in this study was taken using a Photron Ultima APX-RS CCD camera fitted with a bandpass filter such that only light from the $He^{1+}$ line at 468.6nm was imaged at the detector. The plasma conditions are such that at the plasma boundary in MAST, this spectral line is normally well localised to a region around the separatrix.

The spatial resolution of the camera system is ~1.6mm at the tangency plane and frames were recorded at a rate of 500Hz. The camera exposure time was in the range 1.6-2ms in order to record frames that are averaged over turbulence-driven fluctuations in the plasma boundary. The camera location, orientation and imaging properties were determined by fitting the location of unique features in the images that correspond to known locations inside the MAST vessel and using a perspective transformation to project these features on the image. This information was input into an optical ray tracing code to ascertain the path of each camera sightline through the machine, calculate where a ray is tangent to axi-symmetric flux surfaces and locate where it intersects the machine's internal components. It is assumed that most of the light detected at a given pixel originates close to where the pixel's line of sight is tangent. This allows a position in 3D space to be assigned to each pixel in the image for comparison with modelling. Images acquired from the X-point viewing camera before and after the application of an n=6 RMP are shown in Figure 3. The data show that the shape of the plasma boundary is significantly deformed due to the presence of RMPs, as predicted by vacuum modelling calculations.

## 2.1 Lobe Characteristics in H-mode

The reduction in turbulence-driven fluctuations in the plasma boundary due to an edge transport barrier, which is one of the main characteristics of H-mode, gives rise to a sharp plasma boundary in the camera data, as shown in Figure 3. As a result of the sharper definition of the plasma boundary in the camera data, the data collected from H-mode discharges are more straightforward to analyse with image analysis techniques. The data are analysed using the information of the unperturbed magnetic equilibrium, calculated using the EFIT [16] equilibrium reconstruction code, and the location of where each camera pixel line of sight is tangential, to calculate the equilibrium flux at each pixel. In Figure 3, the shape of the plasma boundary (at $\psi_n$ = 1.0) is overlaid on the images in order to determine the apparent deviation of the plasma shape from the equilibrium. It is assumed in the analysis of the camera data that the distribution of the light detected by the camera is due to the deviation of magnetic field lines due to the presence of RMPs, and therefore that the appearance of the lobes in the camera data is mostly due to changes in the magnetic topology.

All of the data collected in this study were recorded from lower single-null plasmas with 600kA plasma current. In the H-mode phase of the discharges, the core electron density was $n_e$ ~ $3.5\times10^{19}m^{-3}$ (resulting in a Greenwald fraction, $I_p/\pi a^2$, of ~0.51) and the core electron temperature was $T_e$ ~ 1keV. The L-mode data used are from the same pulses as the H-mode data, prior to the L-H transition, where the core electron density and temperature were lower compared with the H-mode phase, $2.5\times10^{19}m^{-3}$ (Greenwald fraction ~0.37) and 800eV respectively. The edge safety factor, $q_{95}$, was ~2.8 and the toroidal field at 0.8m was 0.55T in these discharges. In the L-mode phase of the discharges, 2.0MW of auxiliary heating was used via neutral beam injection, rising to 3.6MW in the H-mode phase. Camera data recorded during inter-ELM phases was used in the analysis, only retaining frames from 20-80% of the ELM cycle to remove the effect of ELMs on the data, which have a significant effect on the appearance of the plasma boundary.

**2.1.1 Lobe Spatial Extent**

The apparent size of a lobe is measured by examining the variation of the $He^{1+}$ light emitted from the plasma as a function of the distance from the separatrix, at a given poloidal angle. The camera data is interpolated onto a regular mesh of poloidal angle and radial distance from the last closed flux surface for analysis purposes. An example of this mapping applied to the camera data is shown in Figure 5 in the case of an n=6 RMP, and Figure 6 for an n=4 RMP. The poloidal field coils, and coil supports, which are mounted within the vacuum vessel, which are viewed by the camera introduce a source of reflections and make the background highly non-uniform. In order to reduce the influence of reflections on the analysed data, only lobes that are not close to the poloidal field coils in the camera data are analysed. In the analysis presented here, the lobes highlighted in Figure 7 have been analysed, unless otherwise stated. The lobes furthest from the X-point on the low-field side of the plasma were used in this analysis. This lobe was used as the radial decay of light along this lobe is steepest, due to the reduced flux expansion in this region, so the measurements are less sensitive to density and temperature fluctuations in the far SOL.

The variation of lobe size with RMP field amplitude was investigated by varying the current in the RMP coils in similar discharges, with an n=6 perturbation applied. The size of a lobe was defined to be the distance over which the light intensity falls to 1/3 of its peak (separatrix) value. The analysis is repeated for camera frames prior to the RMP being applied to determine the same light intensity decay length without RMPs. The lobe lengths presented in this paper are the light intensity decay length with RMPs applied minus the equivalent decay length with no RMP applied. Figure 7 shows data from four discharges that the radial extent of a lobe increases linearly with the current in the RMP coils. Furthermore, repeating the analysis with the criterion that the lobe size is the distance over which the light intensity falls to 1/4 of its peak value, shows a similar linear trend. It was observed that the poloidal extent of the lobes was insensitive to the current in the RMP coils. The full width at half maximum of one of the lobes, due to an n=6 RMP, was found to be 0.2±0.1 degrees, corresponding to an arc length of ~3.7mm. This width was not found to vary as the RMP current was increased from 2.4-5.6kAt, within measurement uncertainties. The threshold RMP current at which the lobes appear is determined from the data in two ways. A simple estimate of the threshold can be determined by noting the value of the RMP current when the lobes first become visible in the camera data, which is shown as a vertical gray bar in Figure 7. The second method is to use the linear relationship between the lobe size and RMP current to calculate the threshold. Both of these methods yield similar values of the threshold RMP current, approximately 0.5kAt. The peak signal:noise ratio within the lobes is 25:1, falling to ~8:1 where the length of the lobes are measured, so these results are not significantly affected by the detection limit of the camera.

Data was collected of the variation in lobe size with RMP current with n=4 perturbations applied, as shown in Figure 8. As before, discharges with similar plasmas were run, where the amplitude of the RMP was varied. For the two lobes analysed, both have a linear relationship between radial extent and RMP coil current. Using the scaling data to determine the RMP current threshold for lobes to form, it is found that data from both lobes yield a similar threshold, approximately 0.7kAt. This threshold is in agreement with an approximate threshold found by noting the RMP current when the lobes are first visible in the camera data.

### 2.1.2 Relation to Change in ELM Frequency

Previous detailed studies of the effects of RMPs on ELMs in MAST have shown that there exists a linear relationship between the ELM frequency and the RMP current above a threshold [17]. The data presented in the previous section also suggest there exists a linear relationship between the lobe extent and the RMP current beyond a threshold. A comparison of the ELM frequency and apparent lobe length due to n=4 and n=6 RMPs is shown in Figure 10. The data shows a linear relationship, with an offset suggesting that if there is a relationship between lobes and ELM mitigation, there may exist a minimum lobe length for ELM mitigation, which varies with the toroidal mode number of the RMP.

### 2.1.3 Effect of Rotation Braking on Lobes

The application of RMPs in single-null plasmas in MAST results in braking of the toroidal rotation [17], which reduces the plasma screening of the applied RMP [18]. The effect of rotation braking on lobe length was found by analysing images during the H-mode phase of a discharge at the start of the flat-top of the RMP current where an n=6 RMP is applied. Profiles of plasma toroidal velocity were measured using charge-exchange recombination spectroscopy

(CXRS) at the horizontal mid-plane. In this discharge, the core toroidal rotation breaks from an initial value of 20kms$^{-1}$ before the RMP is applied, to 5kms$^{-1}$ some 30-50ms after the onset of a flat-top of the RMP current. The change in the edge rotation is more modest, 9kms$^{-1}$ at the start of the RMP flat-top, falling to a saturated value of 5kms$^{-1}$. The value of the edge rotation was taken at the location where the normalised flux, $\psi_n$ = 0.95. During the period of interest, the plasma line-average density reduces from 3.4×10$^{19}$m$^{-3}$ to 3.0×10$^{19}$m$^{-3}$. To determine if toroidal rotation plays a role in setting the lobe extent, the variation of lobe extent with RMP coil current was measured at the start of the flat-top in the RMP coil current, when the toroidal rotation is decelerating, and 30-50ms after the start of the flat-top, after the toroidal rotation has stopped decelerating. The data in Figure 9 shows that lobes at the start of the RMP current flat-top appear smaller than after the plasma rotation has decreased, suggesting that greater toroidal rotation results in smaller apparent lobe length.

The relationship between apparent lobe length and edge toroidal velocity was investigated by analysing camera frames taken between the start of the start of the flat-top in RMP coil current and when the toroidal rotation stops decelerating. The data, shown in Figure 11, suggests that as the edge plasma decelerates, the lobes appear larger. The apparent increase in lobe length as the toroidal rotation is relatively modest, a 6mm increase in length for 4km/s reduction in rotation, compared with a measurement uncertainty of ±2mm in the camera data. The data suggests that the apparent lobe length varies linearly with edge toroidal rotation, but the comparatively large experimental uncertainties.

## 2.2 Lobe Characteristics in L-mode

It has been observed that the application of RMPs in the L-mode phase of a discharge can result in the formation of X-point lobe structures, Figure 19. Due to the ballooning nature of edge turbulence, which is greatest on the low-field side of the plasma, the lobe structures are clearer on the high-field side compared with the high-field side. However, the lobes on the low-field side are discernible from the background plasma.

The variation of lobe size with RMP coil current was investigated by measuring the length of a lobe, indicated in Figure 12, during a pulse where the RMP current was ramped up during the L-mode phase of a discharge. Due to the limited quantity of data available of L-mode discharges with RMPs, it was not possible to use data from similar discharges. Due to the intermittency of the plasma on the low-field side, large RMP coil currents, greater than 4kAt, are required in order to clearly discern the lobes. The electron density and temperature at the plasma boundary are approximately $6\times10^{18}m^{-3}$ and 20eV respectively, compared with $2\times10^{19}m^{-3}$ and 50eV in the H-mode phase, measured with Thomson scattering. The difference in plasma conditions between L-mode and H-mode is likely to significantly affect the location of the $He^{1+}$ light emitting region in the plasma, thereby affecting the visible extent of the lobes. It is therefore difficult to make direct comparisons between the L-mode and H-mode data of lobe lengths. The toroidal velocity at the plasma boundary measured by CXRS was approximately 8km/s, although there is insufficient data on the time evolution of the edge rotation, and possible screening of the RMP. The apparent lobe length varies linearly with the RMP coil current, as was observed in H-mode, but the threshold RMP current for lobes to form appears to be lower in L-mode.

## 3. Comparison with Modelling

The camera data has been compared with results from vacuum field line tracing calculations using the ERGOS code [14] with different degrees of sophistication. Verification of the lobe poloidal locations and their extent was carried out by launching field lines from the camera tangency plane and recording the minimum equilibrium $\psi_n$ encountered by the field lines, $\psi_{n,min}$. Comparison between the experimental data and the $\psi_{n,min} = 1.0$ surface is shown in Figure 13. The data suggests that ERGOS can accurately predict the number, location and poloidal extent of the lobes, but over-estimates their radial extent.

Detailed comparison of the camera data with ERGOS calculations was carried out by calculating synthetic camera images. A model for the $He^{1+}$ light emission in the plasma is used where it is assumed that the line emission strength is constant along a flux surface. Using camera data from a similar H-mode discharge with no RMP, the light emission as a function of equilibrium normalised flux is determined by using trial functions for the light distribution. From the light distribution, a synthetic image is calculated by line integrating through the light emission field, to compare with the camera data. Comparison is carried out by plotting the signal from the camera along a line crossing the separatix at the low-field side. Comparison between camera data and ray tracing calculations using a $He^{1+}$ light distribution that gives a good fit to the data is shown in Figure 14. Synthetic images were calculated using the equilibrium flux surfaces, and in the presence of the intrinsic error field (due to imperfections in the poloidal field coils and/or their placement) and error field correction (an n=2 field) using ERGOS and specifying the light distribution as a function of $\psi_{n,min}$. The results of the ray tracing calculations are only slightly modified by accounting for the intrinsic error field and error field correction, showing they do not significantly modify the camera data. This $He^{1+}$ light distribution can then be used to simulate how X-point lobes should appear in the camera data.

Synthetic images of X-point lobes were calculated using ERGOS to trace field lines in the view of the camera in the presence of the equilibrium, intrinsic error, error field correction and n=6 RMP fields. As field lines are traced in ERGOS, $\psi_{n,min}$ and the average equilibrium normalised flux along the field line, $\psi_{n,avg}$, are recorded. The $He^{1+}$ light distribution obtained with no RMP applied was used to calculate synthetic camera data, using $\psi_{n,min}$ and $\psi_{n,avg}$ as magnetic co-ordinates, as shown in Figure 15. Qualitative comparison between the experimental and synthetic camera data suggests that the radial extent of the lobes predicted by ERGOS is over-estimated, irrespective of the whether $\psi_{n,min}$ or $\psi_{n,avg}$ as a basis for the synthetic images.

The effects of plasma screening of the applied RMP have been investigated by using the approach described in [15]. This technique calculates the current required to completely screen an applied RMP over a given range of rational flux surfaces. A spectrum plot of an n=6 RMP with and without the presence of screening currents is shown in Figure 16. Previous results showing the effect of screening currents on lobes that form at divertor footprints due to RMPs [15] suggests that the presence of screening currents reduces the size of the lobes. The effects of screening currents on X-point lobes was investigated by calculating simulated camera data, with the same $He^{1+}$ light distribution as used previously, with screening currents applied over a range of rational flux surfaces. Figure 17 shows simulated camera data in the presence of screening currents applied up to the surfaces corresponding to $\sqrt{\psi_n} = 0.93$ ($\psi_n = 0.86$, m = 10) and $\sqrt{\psi_n} = 0.98$ ($\psi_n = 0.96$, m = 18). The best qualitative match between the simulated and experimental data is found when screening currents are applied up to the $\sqrt{\psi_n} = 0.98$ surface; i.e. the RMP field penetrates 2% in normalised radius. Simulated camera data has been calculated using results based on MARS-F simulations, a single-fluid MHD code that self-consistently calculates the plasma response to the applied RMP. The simulation results also in good qualitative agreement with ERGOS calculations with screening currents applied up to $\sqrt{\psi_n} = 0.98$, as shown in Figure 18. In using the $He^{1+}$ light distribution from the plasma without RMPs applied to calculate the synthetic camera data, this analysis assumes that the plasma properties, as a function of field line co-ordinate, are not significantly affected by the presence of RMPs.

## 4. Summary and Discussion

Experiments have been performed on MAST applying n = 4, 6 resonant magnetic perturbations (RMPs) to lower single null plasmas. It has been observed that the application of RMPs results in the formation of lobe structures in the vicinity of the X-point. These lobe structures have been observed using a CCD camera fitted with a bandpass filter, to record $He^{1+}$ light from the plasma boundary in L-mode and H-mode discharges. The properties of these lobes structures, and their variation with the toroidal mode number and amplitude of the applied RMP has been investigated. It has been observed that the radial extent of the lobes have been observed to vary linearly with the current in the RMP coils both in L-mode and H-mode plasmas. It has been observed that this linear relationship is not dependent on the precise definition of where the edge of a lobe lies. The plasma toroidal rotation has been found to affect the apparent extent of the lobes, where higher toroidal rotation results in smaller lobe structures, although the observed effect is weak. This could be a direct observation of the relationship between plasma rotation and screening, where higher rotation velocities result in greater screening and hence smaller lobe structures.

The camera data of the lobes was compared with modelling based on ERGOS calculations of the magnetic structure of the lobes, together with knowledge of the field lines emitting $He^{1+}$ light, and a ray tracing code. It has been found that vacuum modelling can accurately predict the number and poloidal location of the lobe structures, but over-estimate their radial extent. Taking into account the region of the plasma emitting the $He^{1+}$ light detected by the camera and the camera viewing geometry does not reconcile this difference. The effects of plasma screening of the applied RMP were simulated using an ad-hoc screening model in the ERGOS calculations which apply screening currents to a given range of rational flux surfaces to completely null out the applied RMP field at these surfaces. Applying screening currents over 98% of the plasma significantly improves the agreement between the modelling results and experimental data.

Areas to develop in future work include using a more sophisticated analysis of the camera data, including mapping the camera data in terms of ($\psi_n$-$\psi_{n,boundary}$) and a homoclinic coordinate in the camera tangency plane, which is constant along a field line and equal to the toroidal angle at the outboard mid-plane [19]. Mapping of experimental data in this coordinate system is shown in Figure 20. The lobes appear to be regularly spaced along the x-axis and of approximately equal height, which allows for quantitative analysis of lobe structures on the high-field side, and lobe structures within the confined plasma, appearing as "holes" within the unperturbed separatrix in Figure 1. The effects of impurity transport on the camera measurements will be investigated by imaging the lobes in of $C^{2+}$ light (465nm). In future experiments, measurements of lobe structures in double-null plasmas and lobe structures on the high-field side will be taken, by imaging the region in the vicinity of the upper X-point.

**Acknowledgement**

This work was part funded by the RCUK Energy Programme [grant number EP/I501045] and the European Communities under the Contract of Association between EURATOM and CCFE. To obtain further information on the data and models underlying this paper please contact PublicationsManager@ccfe.ac.uk. The views and opinions expressed herein do not necessarily reflect those of the European Commission. Work of P. Cahyna was funded by the Grant Agency of the Czech Republic under grant P205/11/2341

# Figures

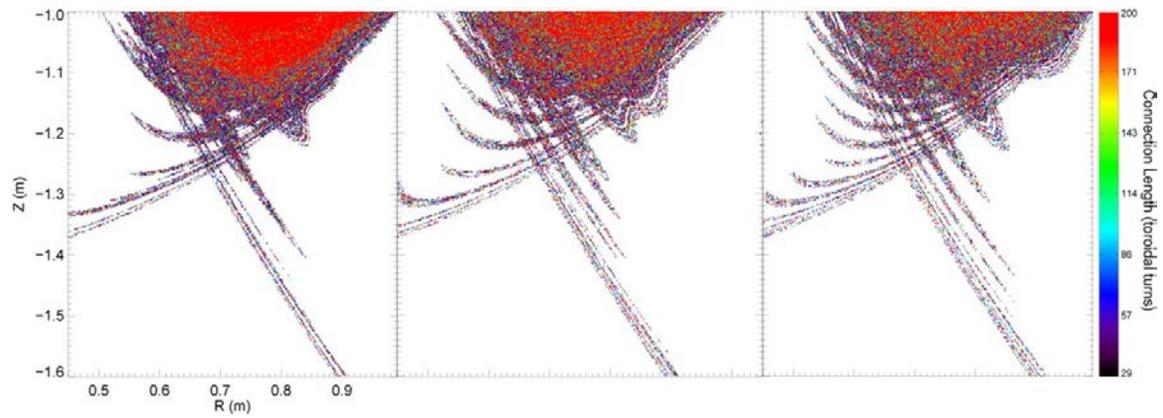

**Figure 1: Results of field line tracing calculations showing the structure of X-point lobes due to n=3 (left), n=4 (centre) and n=6 (right) RMPs.**

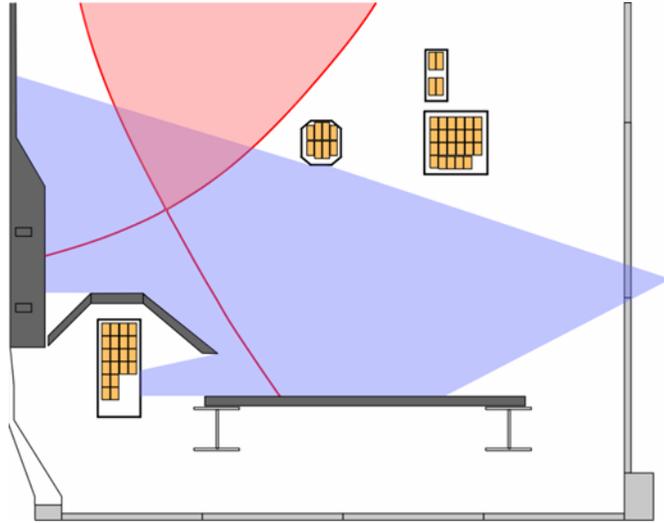

**Figure 2: Schematic view of MAST and the divertor camera viewing geometry (blue).**

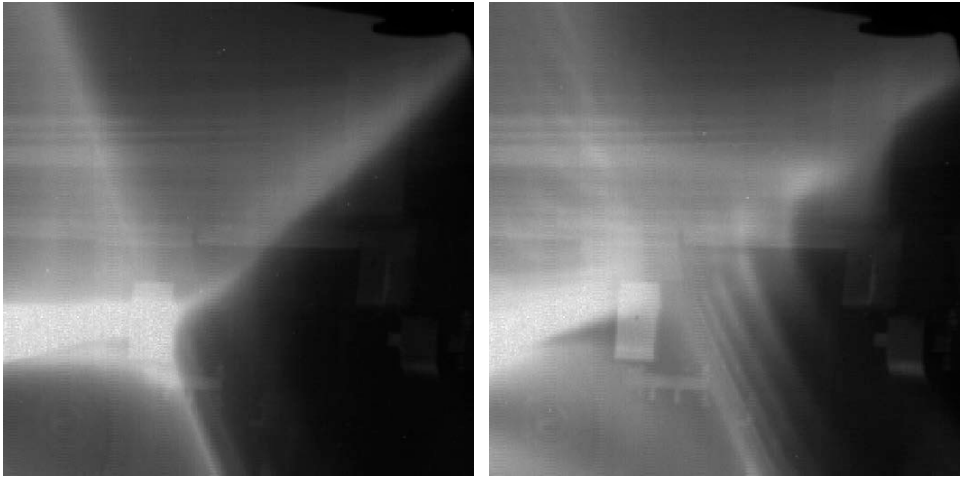

**Figure 3:** The X-point as viewed from the camera in He$^{1+}$ light before (left) and after (right) the application of an n=6 RMP.

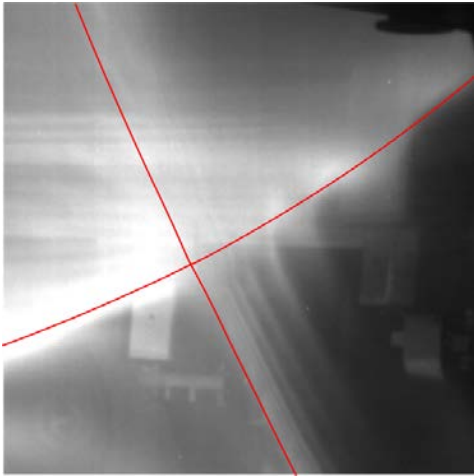

**Figure 4: Raw image of lobes due to an n=6 RMP with the equilibrium separatrix from EFIT superimposed.**

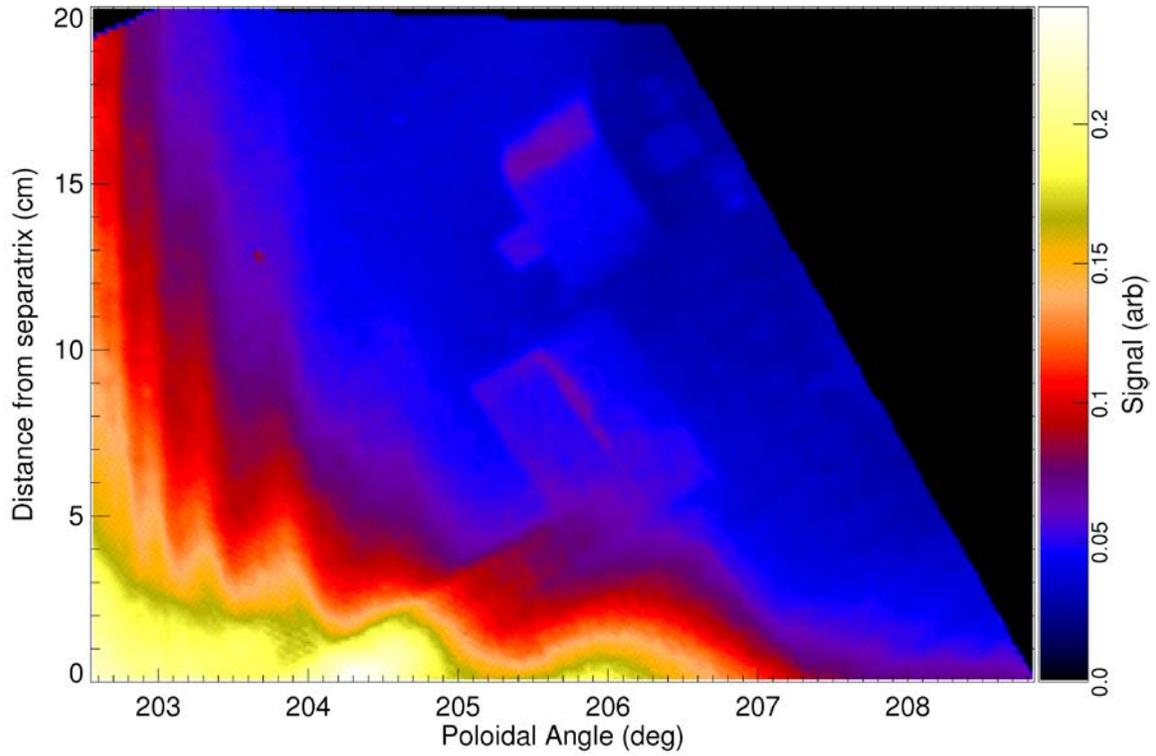

**Figure 5: A camera frame of X-point lobes on the low-field side due to a 4.0kAt n=6 perturbation, plotted in terms of poloidal angle on the ordinate and radial distance from the unperturbed separatrix on the abscissa.**

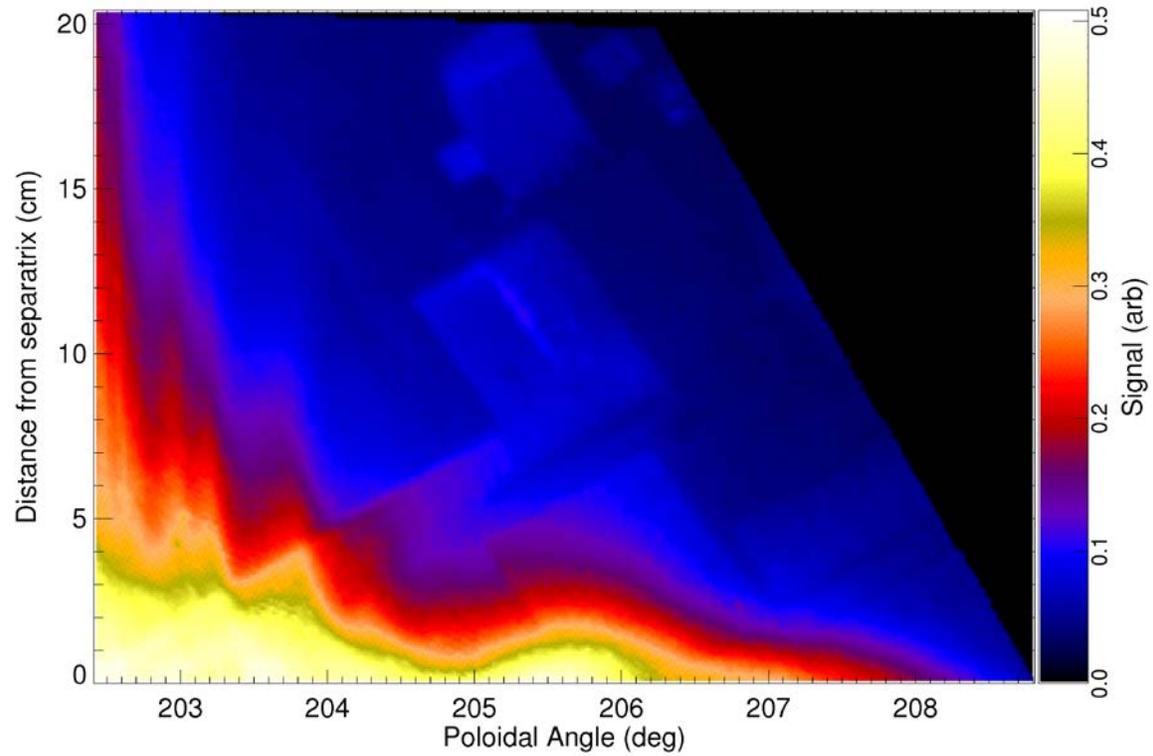

**Figure 6: A camera frame of X-point lobes on the low-field side due to a 5.0kAt n=4 perturbation, plotted in terms of poloidal angle on the ordinate and radial distance from the unperturbed separatrix on the abscissa.**

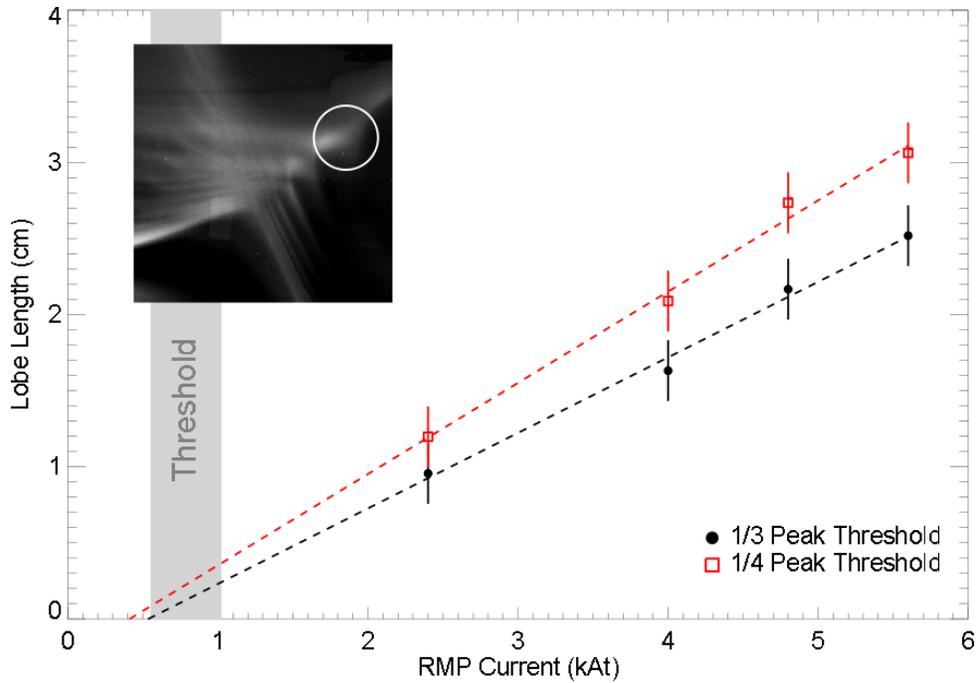

Figure 7: Scaling of apparent lobe extent due to an n=6 RMP with current in RMP coils and comparison between different definitions of where the edge of a lobe lies. The lobe under analysis is indicated in the upper left figure. The RMP coil current when the lobes appear to form in the camera data is indicated by the vertical grey bar.

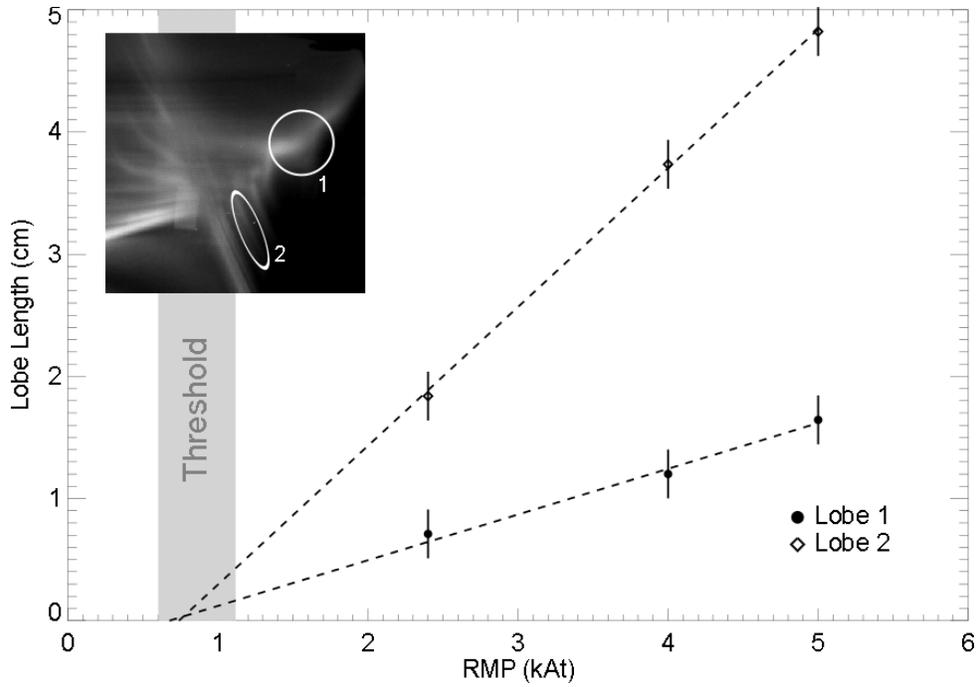

**Figure 8:** Scaling of apparent lobe extent due to an n=4 RMP with current in RMP coils for the two lobes highlighted.

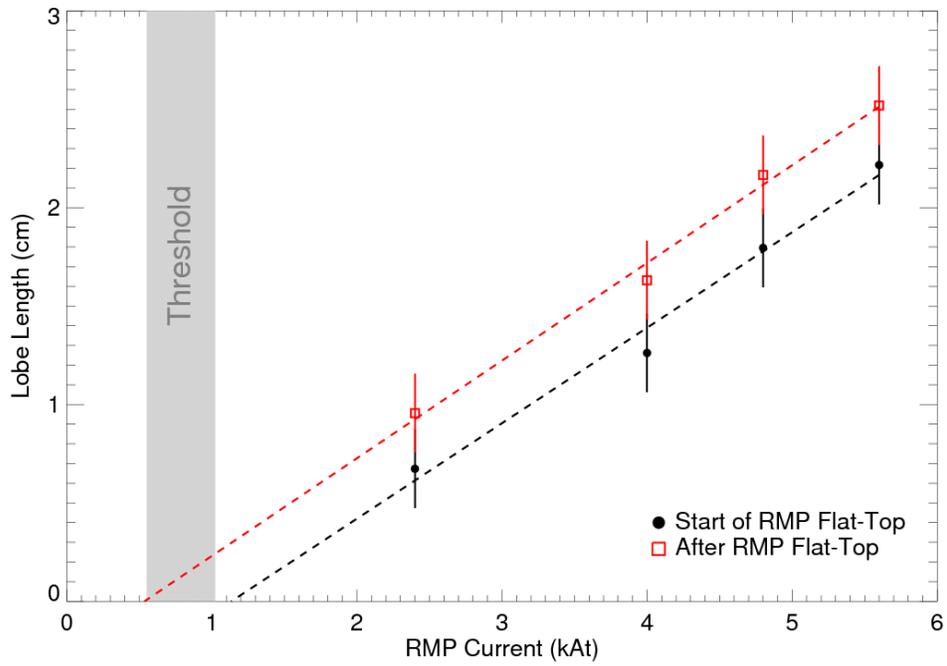

**Figure 9: Comparison of variation of lobe size due to an n=6 RMP with current in RMP coils using data taken at the start of RMP flat-top or 40ms after flat-top onset.**

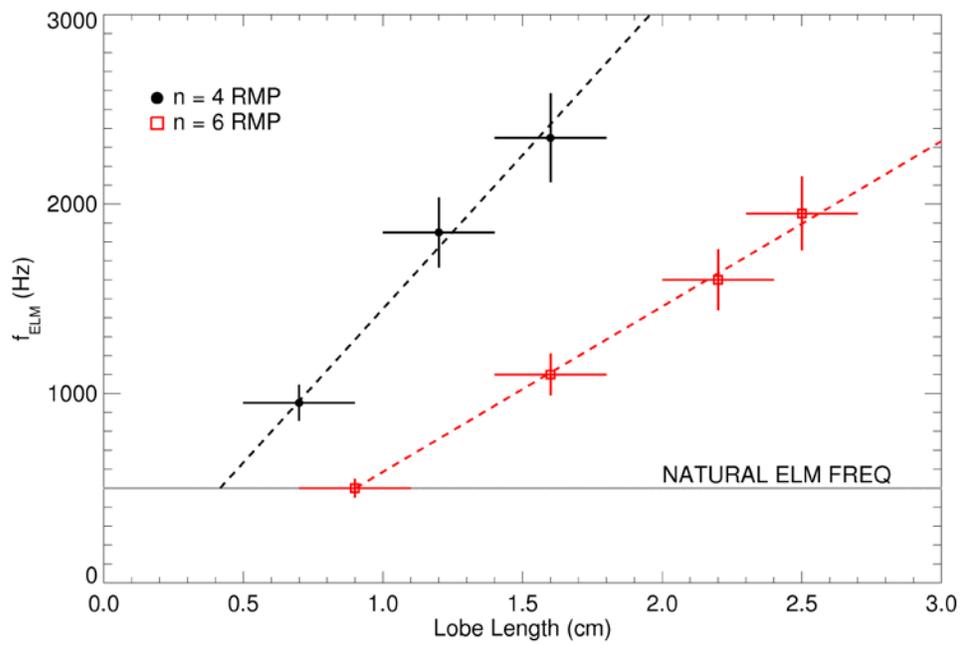

**Figure 10: Comparison of ELM frequency and lobe extent due to n=4 and n=6 RMPs.**

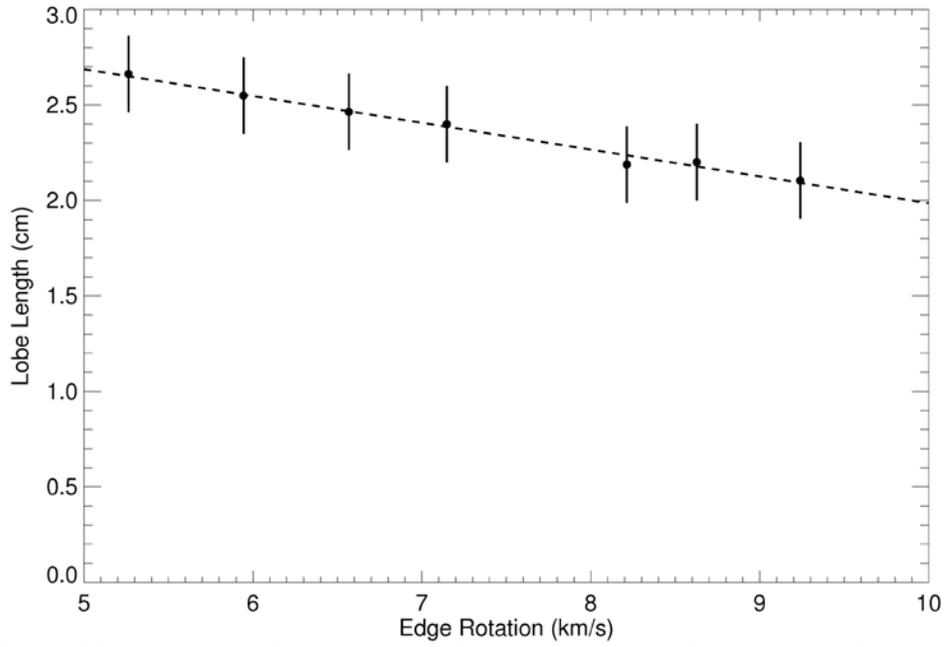
**Figure 11: Length of a lobe compared with edge toroidal rotation as the rotation decelerates after the start of a flat-top in RMP current.**

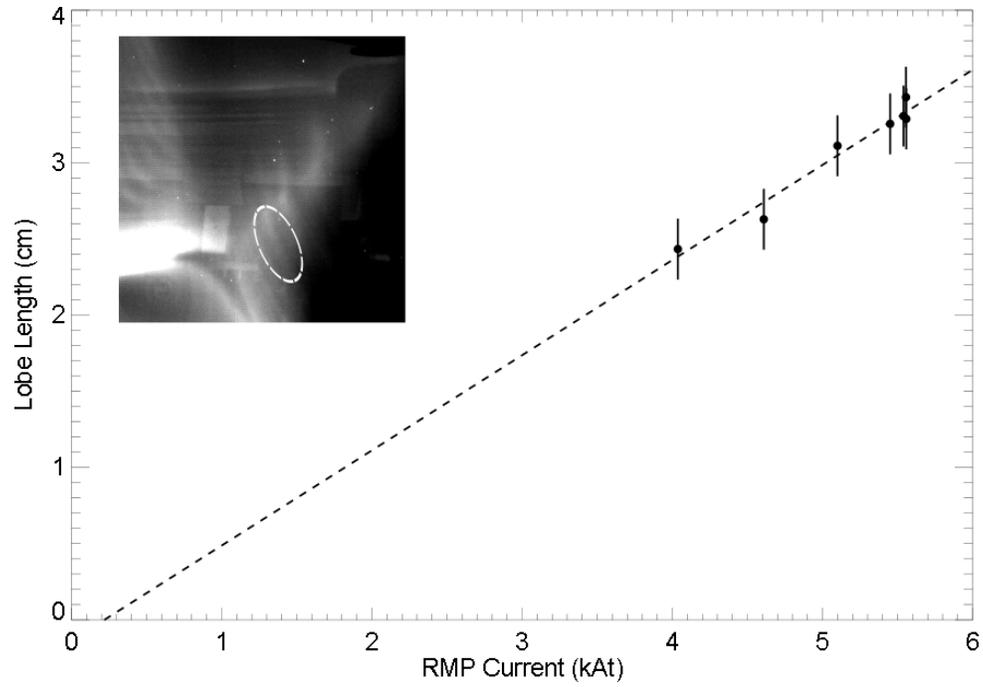
**Figure 12: Scaling of apparent lobe extent due to an n=6 RMP with current in RMP coils in an L-mode plasma.**

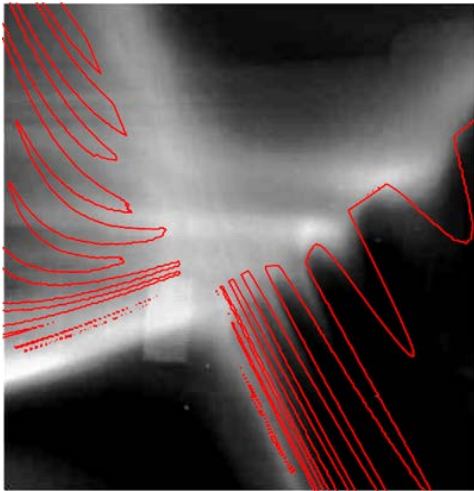
**Figure 13:** Camera data of X-point lobes due to an n=6 RMP with the approximate shape of the plasma boundary predicted by ERGOS, overlaid in red.

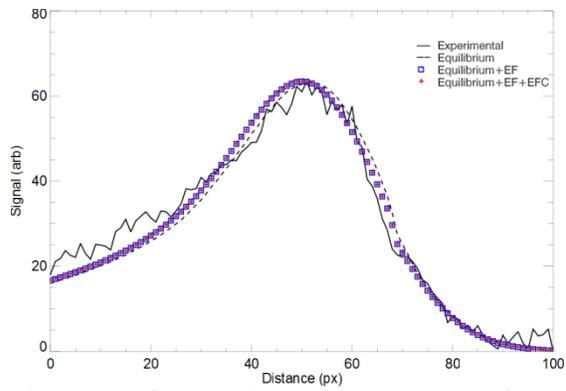

**Figure 14:** Comparison of camera data (black solid line) with simulated data based on ray tracing calculations using data of the 2D equilibrium from EFIT (black dashed line), equilibrium and the intrinsic error field (blue square) and the equilibrium, error field and error field correction (red dot).

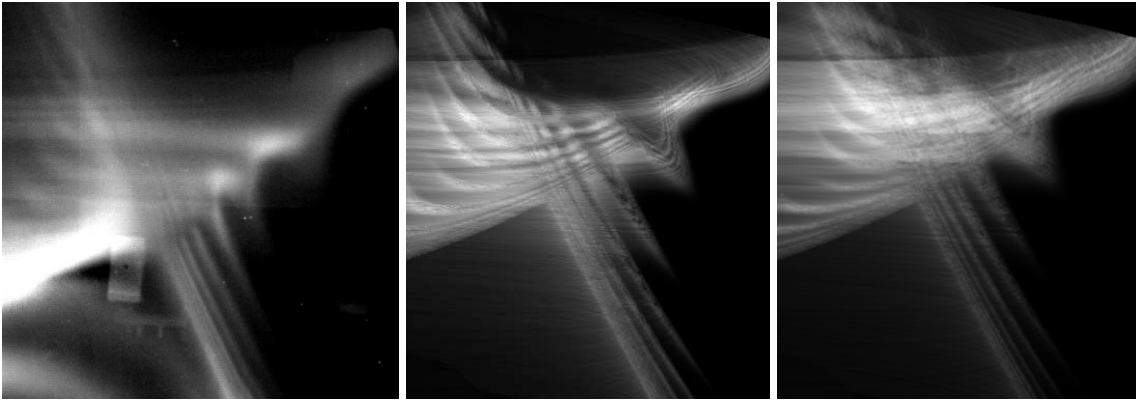

**Figure 15: Left - camera data of X-point lobes due to an n=6 RMP, middle, right - results of ray tracing calculation using ERGOS and a He$^{1+}$ light distribution as a function of $\psi_{n,min}$ (middle) and $\psi_{n,avg}$ (right).**

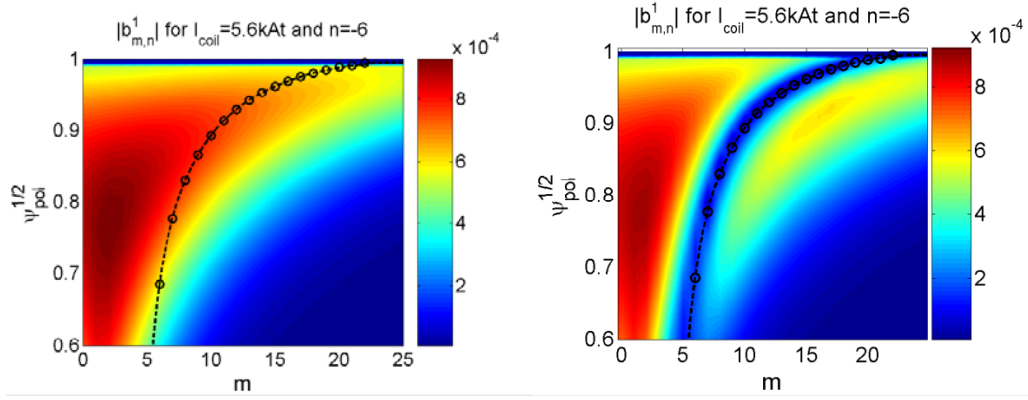
**Figure 16: Spectrum of the applied n=6 RMP from ERGOS without (left) and with (right) screening at rational flux surfaces, with the location of rational flux surfaces overlaid.**

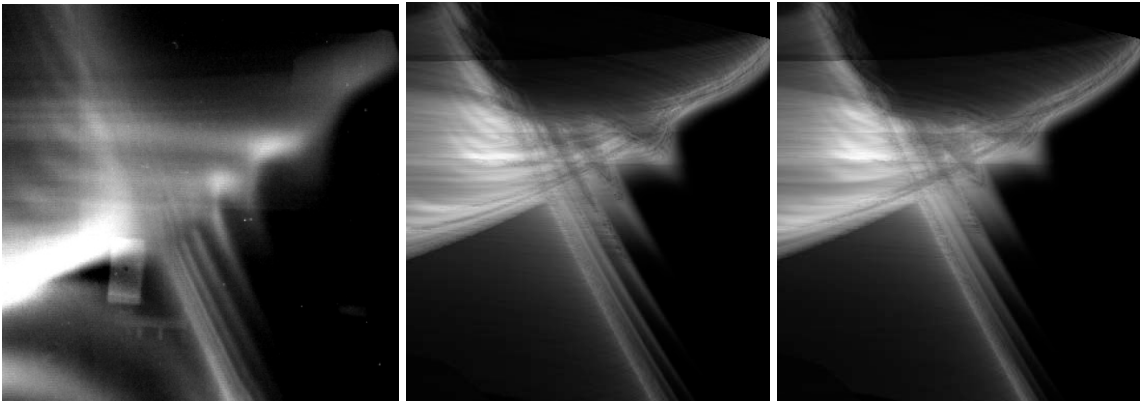

**Figure 17: Left – camera data from Figure 15, right - simulated camera data using ERGOS and screening currents up to $\sqrt{\psi_n} < 0.93$ (centre), $\sqrt{\psi_n} < 0.98$ (right).**

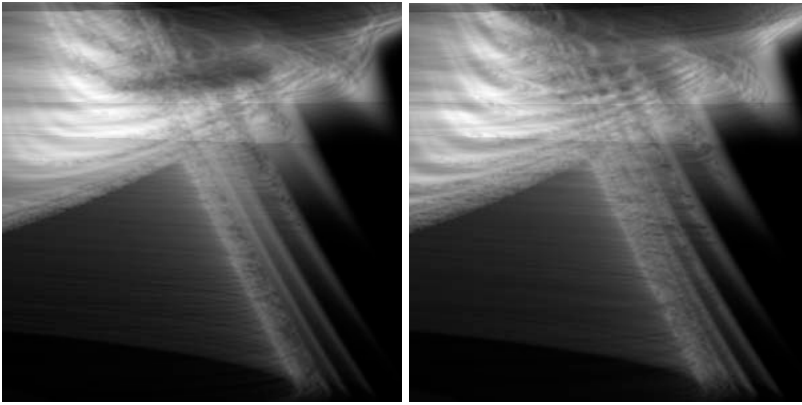
**Figure 18: Simulated camera data using ERGOS and screening currents up to √ψ$_n$ < 0.98 (left) and using MARS-F (right).**

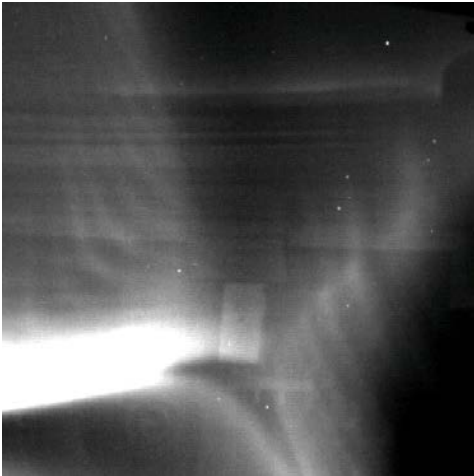

**Figure 19: Appearance of X-point lobes in an L-mode phase. The lobes are most apparent on the high-field side due to the absence of turbulence-driven fluctuations.**

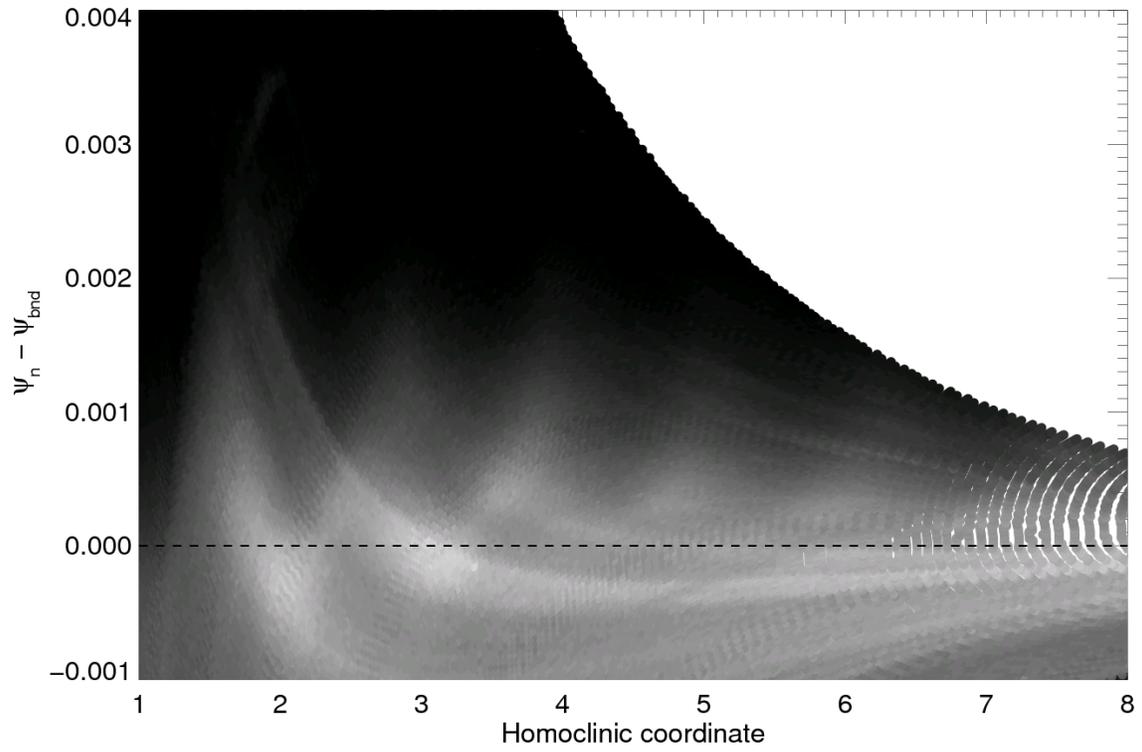

**Figure 20: Mapping of an image of X-point lobes due to an n=6 RMP in terms of flux on the y-axis and the homoclinic coordinate on the x-axis.**